# Diverse Chemistry of Stable Hydronitrogens, and Implications for Planetary and Materials Sciences

Guang-Rui Qian[1,*,#], Haiyang Niu[1,*], Chao-Hao Hu[2,3], Artem R. Oganov[4,5,1,6], Qingfeng Zeng[6], Huai-Ying Zhou[2]

[1]Department of Geosciences, Center for Materials by Design, and Institute for Advanced Computational Science, State University of New York, Stony Brook, NY 11794-2100, USA

[2]Guangxi Key Laboratory of Information Materials, Guilin University of Electronic Technology, Guilin 541004, P. R. China

[3]School of Materials Science and Engineering, Guilin University of Electronic Technology, Guilin 541004, P. R. China

[4]Skolkovo Institute of Science and Technology, Skolkovo Innovation Center, 3 Nobel St., Moscow 143026, Russia

[5]Moscow Institute of Physics and Technology, 9 Institutskiy lane, Dolgoprudny city, Moscow Region 141700, Russia

[6]International Center for Materials Discovery, School of Materials Science and Engineering, Northwestern Polytechnical University, Xi'an 710072, P.R. China

# Abstract

Nitrogen hydrides, e.g., ammonia ($NH_3$), hydrazine ($N_2H_4$) and hydrazoic acid ($HN_3$), are compounds of great fundamental and applied importance. Their high-pressure behavior is important because of their abundance in giant planets and because of the hopes of discovering high-energy-density materials. Here, we have performed a systematic investigation on the structural stability of N-H system in a pressure range up to 800 GPa through evolutionary structure prediction simulations. Surprisingly, we found that high pressure stabilizes a series of previously unreported compounds with peculiar structural and electronic properties, such as the $N_4H$, $N_3H$, $N_2H$ and $NH$ phases composed of nitrogen backbones, the $N_9H_4$ phase containing two dimensional metallic nitrogen planes and novel $N_8H$, $NH_2$, $N_3H_7$, $NH_4$ and $NH_5$ molecular phases. Another surprise is that $NH_3$ becomes thermodynamically unstable above ~460 GPa. We found that high-pressure chemistry is much more diverse than hydrocarbon chemistry at normal conditions, leading to expectations that N-H-O and N-H-O-S systems under pressure are likely to possess richer chemistry than the known organic chemistry. This, in turn, opens a possibility of nitrogen-based life at high pressure. The predicted phase diagram of the N-H system also provides a reference for synthesis of high-energy-density materials.

# Introduction

Hydrogen is the most abundant, and nitrogen is the seventh most abundant element in the universe. Giant planets Uranus and Neptune are predominantly made of H, O, C and N. While the behavior of the H-O[1] and C-O[2] systems under pressure has been investigated in some detail, the N-H system remains largely unexplored. Ammonia ($NH_3$), as an important compound in many branches of science and technology, was firstly reported to exist by Ramsey[3] and Bernal and Mussey[4] in early 1950s, and further discussed by Stevenson and Bundy[5,6]. It is the only stable hydronitrogen at ambient conditions, and exists in a wide range of temperatures and pressures. Recent studies[7–9] revealed that ammonia undergoes a series of phase transitions, including ionic disproportionation and return to non-ionic phase at megabar pressures. Ammonia is considered as a major component of the interiors of giant planets such as Uranus and Neptune under extreme pressure (up to 600 GPa) and temperature (2,000~7,000 K)[10–14]. What has not been properly explored is the full phase stability in the N-H system, including the possibility of decomposition of ammonia; it may well be that, instead of ammonia, very different molecules with different stoichiometries are actually present in planetary interiors.

All nitrogen hydrides, except ammonia, are metastable at ambient pressure. Due to the substantial energy difference between single and triple nitrogen-nitrogen bonds, nitrogen-rich hydronitrogens are potentially superior high-energy-density materials. However, large-scale synthesis of these materials is still problematic. Having a complete phase diagram for the N-H system is necessary for developing synthetic strategies, but such a phase diagram has not been determined. As a result, there is a fundamental interest in investigating the high-pressure behavior and corresponding structural and stability properties of N-H system in both planetary and condensed-matter physics.

Extensive theoretical[15–18] and experimental[19] studies revealed exotic compounds appearing under compression, and exhibiting unique structures and properties different from usual compounds -see previous investigations of NaCl[19], MgO[17], BH[16], H-O[1] and Mg-Si-O[18] systems. Considering the dramatically changed nature of nitrogen[20–22] and the autoionization[7] found in $NH_3$, new hydronitrogen compounds are highly expected to be found.

# Results

**Stoichiometries and structures**

Using the evolutionary algorithm USPEX[15,23–26], we have carried out structure and stoichiometry predictions in order to find all stable compounds (and their stability field) in the N-H system (See Methods). Our calculations confirm that ammonia is the only stable hydronitrogen from ambient pressure to 36 GPa. Above 36 GPa, remarkably, a series of previously unknown compounds become stable, as shown in the pressure-composition phase diagram of the N-H system in Figure 1. The detailed convex hulls at 60, 100, 200, 500, and 800 GPa are presented in Figure 2. It needs to be emphasized that by calculating phonon dispersions, all the newly found compounds in this work are dynamically stable in their corresponding thermodynamically stable pressure zone in the phase diagram. Since zero-point energy can be an factor to affect the relative stability of structures, we have done zero-point energy calculations for $N_2H$, $NH$, $N_3H_7$, $NH_4$ and $NH_7$, and found out that the phase diagram shown in Fig.1 will not change significantly after considering zero-point energy. Therefore, in this work, the phase diagram of N-H is drawn without considering zero-point energy. We would like to leave more accurate phase diagram investigation of N-H at finite temperatures and with considering zero-point energy for further work. These compounds exhibit unusual compositions, peculiar structures and unique properties.

We can classify these thermodynamically stable hydronitrogens compounds that we found into three types (See Table 1). (i) Infinite-chain polymeric hydronitrogens, including $N_4H$, $N_3H$, $N_2H$ and $NH$, with polymeric chains featuring all-nitrogen backbones. (ii) Two-dimensional (2D) metallic $N_9H_4$ phase consisting of 2D nitrogen planes and $NH_4^+$ cations, interestingly, the 2D-nitrogen planes have not been reported in any other nitrogenous compounds before. (iii) Molecular compounds including $N_8H$, $NH_2$, $N_3H_7$, $NH_4$, $NH_5$, and of course $NH_3$. Here, molecular (or molecular ions) compounds are bonded by hydrogen bonds.

**One-dimensional polymeric hydronitrogens**

We have found that, except $N_9H_4$ and $N_8H$, nitrogen-rich hydronitrogens ($N_xH$, $x \geq 1$) are more prone to adopt polymeric structures with N-backbones, The $N_4H$, $N_3H$ and $N_2H$ compounds are predicted to be stable at 51-80 GPa, 42-75 GPa and 60-260 GPa, respectively. The ground state of $N_4H$ has a *Cmc*$2_1$ structure, containing four zigzag nitrogen chains (N-chains) in the unit cell, with pairs of nearest N-chains linked by hydrogen bonds, see Figure 3(a). Here, we use

[ 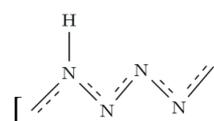 ] to represent the monomeric unit in the polymeric chain of $N_4H$. The delocalized nitrogen-nitrogen bonds run along the zigzag chain, and have the same length of 1.28 Å at 60 GPa. Instead of a zigzag chain, the most stable $N_3H$ structure has space group *P*$2_1$/*c* and is composed of

distorted arm-chair monomers [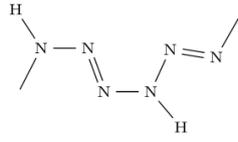], see Figure 3(b). These chains are connected with each other through H-bonds to form a layered structure. The $P2_1/c$ phase of $N_2H$ becomes thermodynamically stable at ~60 GPa, and its structure consists of two [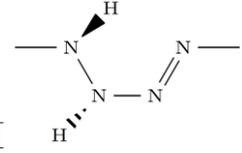] monomers in the unit cell, see Figure 3(c). At 200 GPa, the lengths of single N-N bonds in this polymer are 1.27 and 1.28 Å, and the double N=N bond is slightly shorter (1.24 Å). The small difference hints at a possible bond resonance along the chain. The doubly-bonded nitrogen atoms form weak asymmetric hydrogen bonds with nearby chains. Before the symmetrization of hydrogen bonds occurring at ~280 GPa, $P2_1/c$ -$N_2H$ undergoes a spontaneous decomposition at ~260 GPa. All these polymeric structures are metallic as a result of bond resonance and electronic delocalization along the nitrogen backbone.

With the equal ratio of nitrogen and hydrogen, the NH compound is predicted to be stable in a huge pressure range, from 36 GPa to at least 800 GPa. The $P2_1/c$ structure is more stable than the one predicted in the work of Hu & Zhang[27]. This phase consists of two tetrazene $N_4H_4$ molecules in the unit cell. At 55 GPa, $P2_1/c$ -NH undergoes a phase transition to an ionic structure with $P1$ symmetry. As shown in Figure 3(d), the ionic structure is composed of $N_2H_5^+$ cations arranged in hydrogen bonded layers, alternating with layers of infinite chains [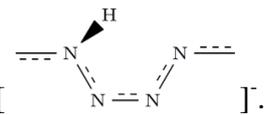]$^-$. The unit cell structure contains 6 NH formula units: $N_2H_5^+$ group and $N4H^-$ from the polymeric chain. At ~180 GPa, all hydrogen bonds become symmetric and the space group raises to $C2$. Both $P1$ and $C2$ are only nominally ionic, because they are metallic and metals have very efficient screening of ionic interactions by the electron gas. Above ~220 GPa, the ionic NH phases become less stable than an $Fdd2$ structure which is made of tetragonal spiral chains, as shown in Figure 3(e). Similar square chains have been reported in group VI elements under pressure, e.g. sulfur-II phase[28] and the $I4_1/amd$ phase of oxygen at pressure around 2 TPa[29]. The $Fdd2$-NH is predicted to be a wide-gap semiconductor (4.8 eV at 400 GPa). In contrast to the strongly localized electrons found in the $I4_1/amd$-oxygen structure with isolated chains, $Fdd2$-NH has asymmetric hydrogen bonds between the square chains. $Fdd2$ transforms to an $Fddd$ structure upon hydrogen bond symmetrization at 460 GPa. For both these orthorhombic phases symmetry breaking leads to two non-equivalent N-N bond lengths in the chain -e.g. 1.25 and 1.34 Å in *Fddd*-NH at 460 GPa. The different lengths of the N-N bonds come from the distortion of the square spirals, caused by their packing and hydrogen bond pattern. The *Fddd*-NH remains stable up to at least 800 GPa.

**Nitrogen hydride with nitrogen planes**

Distracting from the polymeric chain structures, we also discovered an exotic stable nitrogen-rich compound $N_9H_4$. Its structure has *Ccc*2 symmetry, and is composed of negatively-charged 2D nitrogen planes and $NH_4^+$ cations. *Ccc*2-$N_9H_4$ was predicted to be thermodynamically stable in a narrow pressure range 50-60 GPa. As shown in Figure 3(f), the 2D nitrogen plane is a loose structure due to the hexagonal star-shaped holes decorated by 18 additional nitrogen atoms. Parallel stacking of the nitrogen planes creates infinite channels in the perpendicular direction, and $NH_4^+$ cations are located along these channels. The electrons in the plane are delocalized, as a result this compound is metallic with a flat band crossing the Fermi level. (See more details about properties of $N_9H_4$ in Supplementary Information)

**Molecular hydronitrogens**

$N_8H$ is found to be stable around 50 GPa, and adopts a very unusual molecular structure with four pentazole ($N_5H$) and six nitrogen ($N_2$) molecules in the unit cell. (See more details about $N_8H$ structure in Supplementary Information)

Hydrogen-rich hydronitrogens, instead of polymeric structures, have hydrogen-bonded molecular structures. The $NH_4$ phases, containing a higher hydrogen ratio than $NH_3$, are found to be thermodynamically stable above ∼50 GPa, and stable at least up to 800 GPa. At pressures above 50 GPa, $NH_4$ first adopts a host-guest structure of *Pc* symmetry with the structural formula $(NH_3)_2 \cdot H2$. Other host-guest structures, adopting $P2_1$, $C2/c$ and $I4/m$ symmetries, have very close enthalpies to this structure below 80 GPa (See more details about these $NH_4$ structures in Supplementary Information). Accurate fixed composition crystal structure predictions of $NH_4$ show that above 52 GPa, $C2/c$ structure is located in the global minimum of the energy landscape of $NH_4$, and other structures with close enthalpies are structural closely with the $C2/c$ structure. In all host-guest structures, $H_2$ molecules are captured in hydrogen-bonded frameworks formed by $NH_3$ molecules. In the pressure range 85-142 GPa, the ionic $P1$-$NH_4$ phase is more stable than host-guest molecular structures. In the unit cell of this low-symmetry ionic phase, as shown in Figure 4(a), every eighths ammonia molecule reacts with an $H_2$ molecule to form the $NH_4^+$ cation and H⁻ anion. The distance of H⁻ anion and the nearest hydrogen of the NH$_4^+$ cation is 1.13 Å at 100 GPa. Above 142 GPa, the ionic phase undergoes a reentrant transition to the same *C*2/*c* host-guest structure again, thus returning to structures consisting of neutral NH$_3$ and H$_2$ molecules. Hydrogen-bond symmetrization was not observed in all stable NH$_4$ phases up to 800 GPa.

With the 1:1 ratio of H$_2$ and NH$_3$, several NH$_5$ phases are also found to be thermodynamically stable or nearly stable around 55-100 GPa. The ionic *C*2/*c* phase (See Figure 4(b)) has the lowest enthalpy at pressures below 162 GPa. In the unit cell of *C*2/*c*-NH$_5$, there are two [H$_3$N⋯H⋯NH$_3$]$^+$ units and two H⁻ anions. At pressure above ~162 GPa, *C*2/c-NH$_5$ phase transforms into metastable ionic *P*2 and *Ama*2 structures, then adopts a *P*2$_1$/*c* structure containing alternating layers of NH$_3$ and H$_2$ molecules above ~363 GPa. (See more details about NH$_5$ high pressure phases in Supplementary Information.)

At about 140 GPa, a previously unreported remarkable compound with the composition N$_3$H$_7$ is also found to be thermodynamically stable. For N$_3$H$_7$, we have predicted several thermodynamically stable phases with the structural sequence *P*1 → *C*2 → *P*-3*m*1 → *P*2$_1$/*m*-I → *P*2$_1$/*m*-II upon increasing pressure (See Figure 4(c) for the first three structures). At 140-200 GPa, *P*1-N$_3$H$_7$ adopts a stable molecular structure, consisting of one ammonia (NH$_3$) and one hydrazine (N$_2$H$_4$) molecules in the unit cell. At 200 GPa, *P*1 undergoes a spontaneous molecular-to-ionic transition, resulting in a layered *C*2 structure. In this process, ammonia and hydrazine molecules react to form the NH$_2^-$ (amide) anions and N$_2$H$_5^+$ (hydrazinium) cations, respectively. The N$_2$H$_5^+$ ions are in a parallel arrangement and connected by symmetric H-bonds. At 300-380 GPa, complicated ionic N$_3$H$_7$ structure of *P*-3*m*1 symmetry becomes stable. As shown in Figure 4(c), in this unique structure, the trigonal unit cell has two neutral ammonia molecules, one N$^{3-}$ anion, one [N$_2$H$_6$]$^{2+}$ cation and one [ 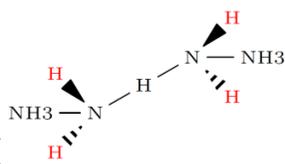 ]$^+$ unit (net formula N$_4$H$_9^+$, the red H symbols indicate that such hydrogen atoms are symmetrically hydrogen-bonded and shared with neighbor N$_4$H$_9^+$ units). This is the only structure with bare nitrogen anions observed among the newly proposed nitrogen hydrides. The nitride anion N$^{3-}$ is surrounded by 12 hydrogen atoms from NH$_3$ molecules and N$_4$H$_9^+$ cation, with distances of 1.32 and 1.38 Å at 380 GPa. Then, at pressure above 380 GPa, the trigonal N$_3$H$_7$ phase will give way to another two *P*2$_1$/*m* type ionic structures, consisting of NH$_2^-$ anions and N$_2$H$_5^+$ cations again. They have different packing patterns from the ionic *C*2 structure (named $P2_1/m$-I and $P2_1/m$-II $N_3H_7$ by stability sequence upon increasing pressure, respectively (See Supplementary Information for more details).

With pressure increasing, our calculation confirmed that $NH_3$, above 36 GPa, undergoes phase transformations from hydrogen-bonded molecular $P2_12_12_1$ structure to layered ionic $Pma2$ and $Pca2_1$ phases, and then returns to $Pnma$ structures consisting of neutral $NH_3$ molecules at very high pressure[7,9]. However, $NH_3$, the only thermodynamically stable hydronitrogen compound at ambient conditions, is surprisingly predicted to decompose into $N_3H_7$ and $NH_4$ at ~460 GPa at zero temperature. For $NH_2$, the dense molecular hydrazine phase was also predicted to be stable and have a $C2/c$ symmetry at ~200-780 GPa, which is consistent with Zhang's work[30]. The $C2/c$ structure of $NH_2$ consists of hydrazine molecules, forming both symmetric and asymmetric hydrogen bonds with each other.

## Discussion

Our theoretical calculations indicate that the N-H system exhibits rich chemistry under pressure. The infinite long-chain polymeric structures are widely found in nitrogen-rich hydronitrogen compounds, and are thermodynamically stable above 42 GPa. They could potentially serve as good high-energy-density and fuel materials due to the substantial energy difference between the single/double and triple nitrogen-nitrogen bonds. The nitrogen backbone follows different patterns as a different hydrogen ratio in compounds. With the "antiseeds" technique (See Methods), we found that metastable nitrogen phases containing zigzag N-chains have competitive enthalpies (~0.03 eV/atom higher at 60 GPa) to the molecular states and the singly bonded cg-N[31] structure at 40-70 GPa, and they are more energetically favorable than arm-chair-shaped and other N-chains (See Fig. S5 in Supplementary Information). A low hydrogen content stabilizes these chains and does not change much of the packing pattern of the chains and the electronic properties of the resonant N-N bonds. The ground states of metastable $N_9H$ and high pressure $N_8H$ phases contain infinite zigzag N-backbones. (See more details about these two compounds in Supplementary Information) With higher hydrogen content, the zigzag N-backbone become unstable in $N_3H$, $N_2H$ and NH phases.

These long-chain polymeric hydronitrogen compounds would be a great alternative to commonly used high-energy-density materials. Compared to pure polymeric nitrogen (cg-N phase), layered $P21/c$ $N_3H$ is stable starting from ~42 GPa, i.e. at pressures lower than the stability pressure of cg-N (> 56 GPa). Hydrazoic acid[32] ($N_3H$) may be an even better precursor for synthesizing long-chain polymers. With hydrazoic acid, the layered $P21/c$ $N_3H$ can be formed at as low as 6.0 GPa (See Table 2). The VC-NEB[33] calculation indicates that the phase transformation from hydrazoic acid to $P2_1/c$ N$_3$H has an energy barrier of ~0.25 eV/atom at 10 GPa, (See Figure 5), and occurs in several stages. In the first stage, some H-bonds between HN$_3$ molecules break making the molecules free to rotate (as shown in Figure 5 from Image-1 to Image-5). After adjusting directions of HN$_3$ molecules (Image-5 to Image-18), metastable short N-chain molecules (Image-21 and Image-27) appear during the transition, new nitrogen-nitrogen bonds appear, eventually leading to infinite polymeric chains (Image-30). The energy barrier of first stage with rotation of the HN$_3$ molecules is around 0.15 eV (from Image-1 to Image-19), and approximately equals to the barrier of the second stage (nitrogen-nitrogen bond formation). The transition should happen easily in liquid hydrazoic acid. Mixture of hydrazine and hydrazoic acid are an alternative precursor, with polymerization estimated to happen at ∼13 GPa (See Table 2).

Isoelectronic to oxygen, (NH) units generally serve as analogs of group VI elements in these polymeric chain structures. Besides the square-spiral chain in high pressure phases found in NH, the monoclinic N$_2$H phase can be considered as an analogue material of sulfur nitride (SN)$_n$[34] or (ON)$_n$[35, 36] polymers. The proposed nitrogen oxides (ONNO)$_n$ chain oligomer also has comparatively strong N=N bonds. The monoclinic N$_2$H phase is a metallic polymer as the Fermi level is crossed by anti-bonding π* bands (See Fig.S4 in Supplementary Information), which is similar to the first known metallic polymer (SN)$_n$[37] as a superconductor with $T_c$ = 0.26 K[38]. All our 1D long-chain hydronitrogen compounds containing delocalized nitrogen bonds are metallic. Our calculations reveal that N$_4$H (at 55 GPa) and N$_2$H (at 60 GPa) are superconductors with $T_c$ = 2.6 and 7.8 K (with the value of μ*=0.13), respectively. In contrast, N$_9$H$_4$ phase is not a superconductor.

Multiple stable stoichiometries also exist in hydrogen-rich hydronitrogens at pressure. These hydronitrogens form molecular crystals at low pressure, and then tend to undergo auto-ionization under moderate compression, except NH$_2$ (See Table 1). The structures of these compounds show various characteristics and are quite different from each other. N$_3$H$_7$, NH$_4$ (and NH$_5$) can be considered as binary NH$_3$ + N$_2$H$_4$ and NH$_3$ + xH$_2$ compounds, respectively. Therefore, in general, high-pressure hydrogen-rich hydronitrogens tend to contain molecules and molecular ions.

It is predicted that hydrogen-rich hydronitrogens remains stable to extremely high pressures, NH$_3$ and NH$_2$ become unstable and decompose (into NH$_4$ and N$_3$H$_7$, or into NH and N$_3$H$_7$) only at 480 and 780 GPa, respectively; and NH$_4$ and N$_3$H$_7$ are thermodynamically stable at least up to 800 GPa. In contrast, methane (CH$_4$) was predicted to dissociate into ethane (C$_2$H$_6$), butane

($C_4H_{10}$), and finally, diamond plus hydrogen at 287 GPa[2].

$NH_4$ and $NH_5$ undergo a molecular⇒ionic⇒molecular phase sequence under pressure, which is very similar to $NH_3$[7]. The auto-ionization process also occurs in $N_3H_7$, which remains in the ionic phase at least up to 800 GPa. In contrast, C-H compounds have non-polar non-ionic structures, and the high energy cost of proton transfer in $H_2O$[7, 39] prevents auto-ionization until extremely high pressure (~1.4 TPa)[39]. Our calculation revealed that the energy cost of proton transfer from $H_2$ to $NH_3$ molecule and from $NH_3$ to $N_2H_4$ molecule is ~0.7 eV and ~1.0 eV, respectively, while it costs ~0.9 eV[40] to form $NH_2^-$ and $NH_4^+$ ions in $NH_3$. Therefore, $NH_3 + xH_2$ compounds would undergo auto-ionization at a lower pressure ($NH_4$ at ~85 GPa and $NH_5$ at ~42 GPa) than pure $NH_3$ (at ~90 GPa). Due to high cost of proton transfer, auto-ionization phenomenon was not observed in any stable $H_2O$-$H_2$ compounds[1]. Calculations show that auto-ionization happens at ~200 GPa in $N_3H_7$, higher in $NH_3$ (90 GPa)[7], due to the higher proton transfer energy cost, and survives up to at least 800 GPa. The pV term in the free energy plays an important role in deterring the phase transition sequence at high pressure. Under pressure, stable $N_3H_7$ and $NH_3$-$xH_2$ host-guest phases are more packing efficient than the summary volume of $NH_3 + N_2H_4$ and $NH_3 + H_2$, respectively. The auto-ionization transition in $N_3H_7$ leads to denser structures and enhances stability of $N_3H_7$ under compression.

## Conclusions

We have extensively explored the nature of hydronitrogen compounds up to ultrahigh pressures. It turns out that unusual compounds, such as $N_8H$, $N_4H$, $N_3H$, $N_9H_4$, $N_2H$, $NH$, $NH_2$, $N_3H_7$, $NH_4$ and $NH_5$ are stable under pressure. These compounds possess intriguing crystal structures and remarkably novel, exotic properties. Three main features can be concluded, 1) the (NH) unit behaves similarly to its isoelectronic analogs, oxygen (also the sulfur) atoms, 2) molecular hydronitrogens are mainly composed of $H_2$, $NH_3$, $N_2H_4$ molecules and corresponding ions, 3) auto-ionization is common in N-H molecular phases due to the low energy cost of the proton transfer between the $H_2$, $NH_3$, $N_2H_4$ molecules.

Our investigation opens ways for designing synths of novel high-energy-density polymeric hydronitrogens. It is clear that starting with metastable precursors (such as $N_2H_4$, $N_3H$) should lower polymerization pressure (compared to the lowest pressure of thermal dynamic polymeration, 42 GPa). We experimented with different mixtures of $N_2H_4$, $N_3H$ and $N_2$ give bulk $N_3H$ or NH compositions. We found that using $N_2$ in the precursor mixture does not give good results. Instead, pure $N_2H_4$ and $N_3H$, or their mixtures can polymerize already at near-ambient conditions. For planetary interiors (where H/N > 1), we expect the presence of N-contianing molecular ions at all pressures above ~55 GPa in $NH_5$. This means a much thicker layer with ionic conductivity than previously thought, which will affect models of planetary magnetic fields (which are generated by convection of electrically conducting layers). High-pressure chemistry of hydronitrogens uncovered here has greater diversity than hydrocarbons. This invites the question whether nitrogen-based (rather than carbon-based) life is possible in the interiors of gas giant planets. This is not impossible -thought at high temperature of planetary interiors lifetime of metastable molecules (essential for life) will be short.

We remind that at normal conditions, the only thermodynamically stable compound of carbon and hydrogen is methane ($CH_4$), all the other hydrocarbons being metastable and kinetically protected by high energy barriers. Here we have uncovered unique structural diversity among THERMODYNAMICALLY STABLE hydronitrogens. N-H bonds are directional covalent bonds (just like C-H), which should also lead to high energy barriers and ubiquitous metastability. If one includes metastable hydronitrogens, and adds other elements (such as O, S, smaller amounts of C), the diversity will most likely exceed the diversity of organic chemistry. This invites the question whether nitrogen-based (rather than carbon-based) life is possible in the interiors of gas giant planets. Briefly, we see the following conditions as necessary for emergence of life: (1) great structural and chemical diversity based on a small number of chemical elements (C-H-O or N-H-O), (2) abundance of metastable compounds with long lifetime, (3) a chemical reaction that can provide energy, (4) reversible reaction for storing/releasing energy (similar to the function of ATP in carbon-based life), (5) a molecule that can be used as information matrix (analogous in its function to DNA). For nitrogen-rich compounds, condition (1) is clearly satisfied. Condition (2) is also likely satisfied at not very high temperatures. Energy source and storage can be related to metastable compounds -e.g. oxidation of hydronitrogens for energy production, and polymerization/depolymerization of hydronitrogens for energy storage. As for condition (5), it is too early to say which N-based molecules could be suitable -the main conditions seem to be 1D-or 2D polymeric nature and aperiodicity. Nitrogen-based life could be possible, but the likelihood of this is highly limited due to high temperatures (up to ∼7000 K) in these planets' interiors, which makes lifetimes of metastable compounds very short. Given the abundance of N, H, O, C in giant planets, and high pressures in their interiors, we expect great chemical diversity there.

# Methods

**Crystal structure prediction**

Crystal structure prediction was performed using the variable-composition evolutionary algorithm USPEX[15, 23–26]. A number of studies illustrate the power of the USPEX method[16, 17,19]. Calculations for the N-H system were performed at various pressures in the wide range of 0-800 GPa.

Given the dramatically changed behavior of nitrogen under pressure and a wide pressure range of our investigation, we performed a number of different types of predictions with USPEX. We ran variable-composition predictions for N-H, N-NH and NH-H systems with up to 32 atoms per unit cell. Given molecular nature of all stable and nearly stable compounds in hydrogen-rich hydronitrogens, we also did structure prediction for the packing of well-defined $NH_3$ and $H_2$ molecules (rather than N and H atoms), by applying the specially designed constrained global optimization algorithm, considering structures with up to 24 molecules (i.e. up to 96 atoms) per primitive unit cell. These calculations were run together in a global coevolutionary search with exchanging good (stable and some metastable) structures between different runs. This coevolutionary method is very efficient and has been implemented on top of the USPEX code. When performing prediction for metastable nitrogen structures containing zigzag N-chains, we applied the antiseeds technique[41], which was adopted to search for all low-enthalpy structures based on zigzag N-chains.

**DFT calculations**

The underlying *ab initio* structural relaxations and electronic structure calculations in UPSEX were carried out using the all electron projector augmented wave (PAW)[42] method as implemented in the VASP code[43]. The plane-wave cutoff energy of 800 eV and dense Gamma-centered k-point meshes with a resolution better than $2\pi \times 0.05$ Å were adopted, and ensured high-quality results. After identifying the most stable compositions and several candidate structures, we relaxed them at numerous pressures in the range of 0-800 GPa with harder PAW potentials, in which the core radius equals 0.42 and 0.58 Å for hydrogen and nitrogen, respectively. An extremely high cutoff energy of 1400 eV was used for these relaxations and calculations of enthalpies of reactions and phase diagram. In addition, phonon dispersions throughout the Brillouin zone were derived using the finite-displacement approach as implemented in the Phonopy code[44]. Superconducting $T_c$ was calculated in QUANTUM ESPRESSO[45], with ultrasoft potentials[46] using 40 Ry plane-wave cutoff energy.

## Acknowledgements

This work was supported by National Science Foundation (EAR-1114313, DMR-1231586), DARPA (Grants No. W31P4Q1210008 and No. W31P4Q1310005) and the Government (No. 14.A12.31.0003) of Russian Federation (Project No. 8512) and Foreign Talents Introduction and Academic Exchange Program (No. B08040). C.-H. Hu thanks the National Basic Research Program of China (973 Program, Grant No. 2014CB643703), National Natural Science Foundation of China under 11164005 and 51372203, Guangxi Natural Science Foundation under 2014GXNSFGA118001 and 2012GXNSFGA060002, and Guangxi Key Laboratory of Information Materials (Grant no. 1210908-215-Z). The authors also acknowledge Purdue University Teragrid and TACC Stampede system (Charge No.: TG-DMR110058) and High Performance Computing Center of NWPU for providing computational resources and technical support for this work.

## Author contributions

G.R.Q. and H.Y.N. contributed equally to this work. A.R.O., C.H.H. and G.R.Q. designed research, G.R.Q., H.Y.N., C.H.H., Q.F.Z. and H.Y.Z. performed simulations, G.R.Q. wrote the coevolution code, G.R.Q. and H.Y.N. analyzed data, G.R.Q., H.Y.N., A.R.O. and C.H.H. wrote the manuscript.

## Additional information

Competing financial interests: The authors declare no competing financial interests.

# Tables

**Table 1** Structure details of stable nitrogen hydrides compounds

| Compounds | Pressure(GPa) | Structure Type | Subunits |
|---|---|---|---|
| $N_8H$ | 50 ~ 54 | molecular | $NH_5 + N_2$ |
| $N_4H$ | 51 ~ 80 | long-chain | 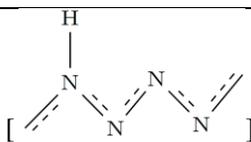 |
| $N_3H$ | 42 ~ 75 | long-chain | 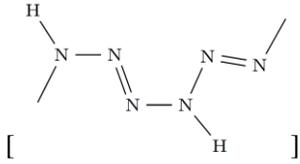 |
| $N_9H_4$ | 50 ~ 60 | two dimensional | $N_x^-$ plane + $NH_4^+$ |
| $N_2H$ | 60 ~ 260 | long-chain | 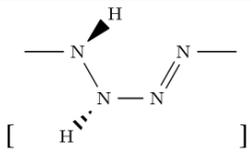 |
| NH | 36 ~ >800 | (a) short chain ($P2_1/c$) | $N_4H_4$ |
|  |  | (b) dimer+long-chain ($P1$, $C2$) | $N_2H_5^+$ + [ 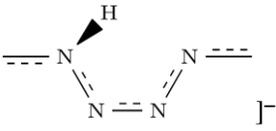 ]$^-$ |
|  |  | (c) long-chain ($Fdd2$, $Fddd$) | tetragonal spiral chain |
| $NH_2$ | 200 ~ 780 | molecular | $N_2H_4$ (hydrazine) |
| $N_3H_7$ | 140 ~ >800 | (a) molecular/ionic | $NH_3 + N_2H_4 / NH_2^- + N_2H_5^+$ |
|  |  | (b) ionic ($P\text{-}3m1$) | $N_3^- + N_2H_6^{2+}$ + [ 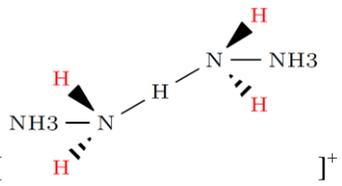 ]$^+$ |
| $NH_3$ | 0 ~ 460 | molecular/ionic | $NH_3 / NH_2^- + NH_4^+$ |
| $NH_4$ | 50 ~ >800 | molecular/ionic | $NH_3 + H_2 / NH_3 + NH_4^+ + H^-$ |
| $NH_5$ | 55 ~ 100 | ionic | $NH_3 + [H_3N \cdots H \cdots NH_3]^+ + H^-$ |

**Table 2** Chemical reactions to synthesis high-energy-density hydronitrogen at ΔH=0

|  | Reaction | Pressure [GPa] | ΔV |
|---|---|---|---|
| $N_3H$ (HA*) | → $N_3H$ (long-chain) | 6.0 | 7.58 |
| $N_2H_4 + N_3H$ (HA) | → 5NH (dimer+long-chain) | 12.1, 12.8** | 10.9, 9.81** |
| $N_2H_4 + N_2(P4_12_12_1)$ | → 4NH (dimer+long-chain) | 32.5 | 6.26 |
| $N_2H_4 + 5N_2(P4_12_12_1)$ | → $4N_3H$ (long-chain) | 37.3 | 18.5 |

* HA shorts for Hydrazoic Acid

** With $C2$ and $P2_1$-$N_2H_4$ phases, respectively

**Figure captions**

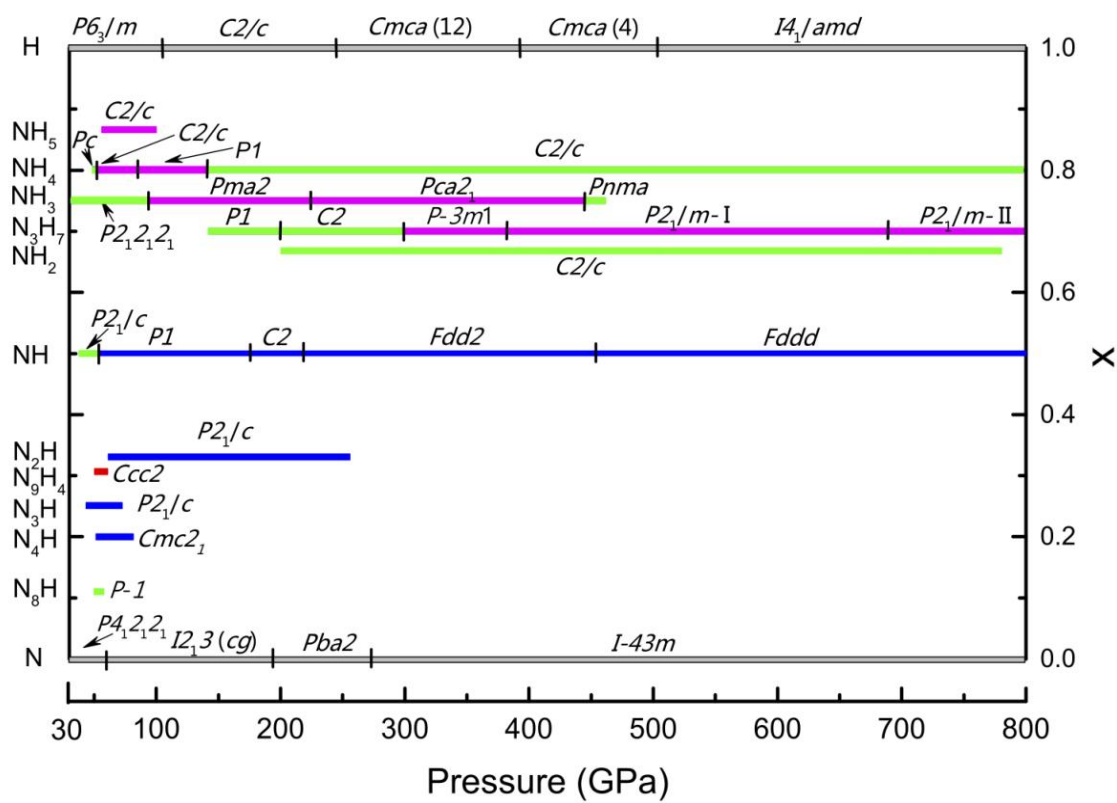

**Figure 1** Phase diagram for N-H system from 30-800 GPa. For the hydronitrogen structures, the phases with blue color indicate infinite nitrogen chain structures. The green and pink phases indicates molecule and molecular ionic structures, respectively. The red color indicates the 2D-plane $N_9H_4$ phase.

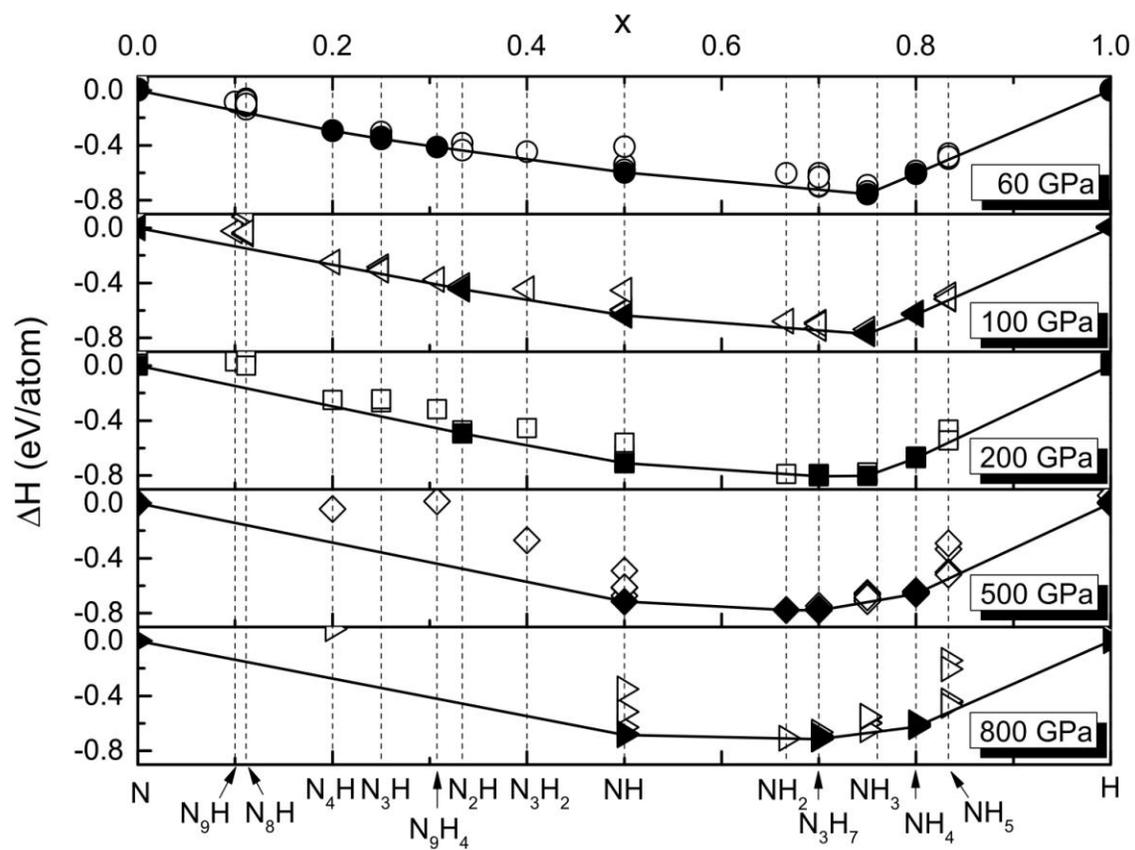

**Figure 2** Convex hull for nitrogen hydride system at 60, 100, 200, 500, 800 GPa. The solid and hollow symbols indicate stable and metastable phases, respectively.

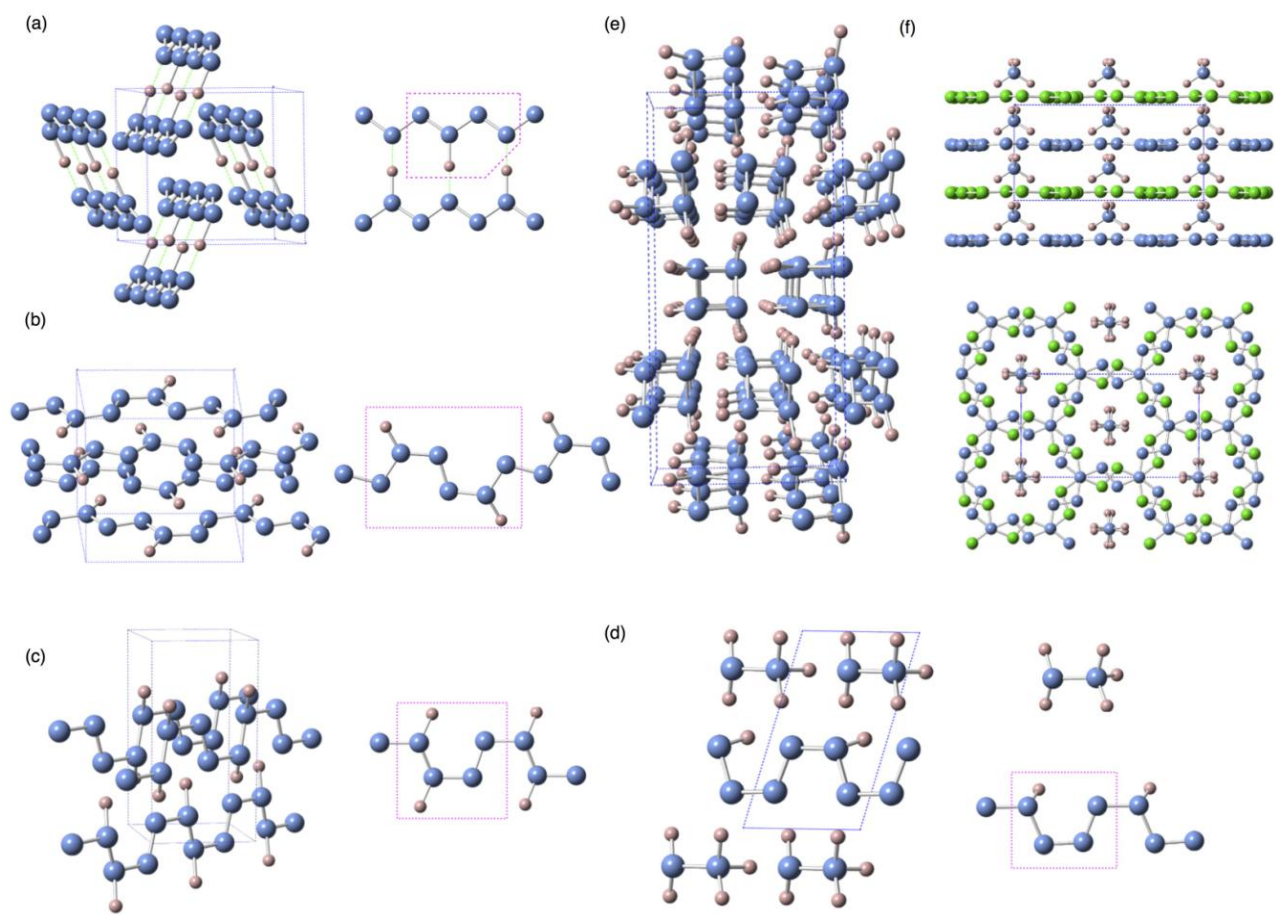

**Figure 3** The proposed structures for $N_4H$, $N_3H$, and $N_2H$ and NH, $N_9H_4$. The small pink spheres indicate hydrogen atoms and the blue large spheres are nitrogen atoms. The structures in the pink box are the corresponding monomeric units. (a) $Cmc2_1$-$N_4H$ structure. The structure is composed of one dimensional zigzag-shaped N-chains. Every two chains are engaged though asymmetric hydrogen bone, and crosswise packed. (b) Layered $P2_1/c$-$N_3H$ structure containing distorted arm-chair-shaped chain. (c) $P2_1/c$-$N_2H$ structure composed of parallel one dimensional arm-chair-shaped $N_2H$ chains. (d) $P1$-NH structure. Its structure consists of $N_2H_5^+$ ion and negatively charged arm-chair-shaped chain layers. It will transform to $C2$ phase at 180 GPa, due to the symmetrization of the hydrogen bonds between $N_2H_5^+$ ions and between chains. (e) The $Fdd2$-NH structure consists of quare-spiral-like chains. The length of nitrogen bond along [0 1 0] and [0 0 1] plane are not the same. (f) Top view and side view of $Ccc2$-$N_9H_4$. The small pink spheres indicate hydrogen atoms and the blue and green large spheres are nitrogen atoms at different layers.

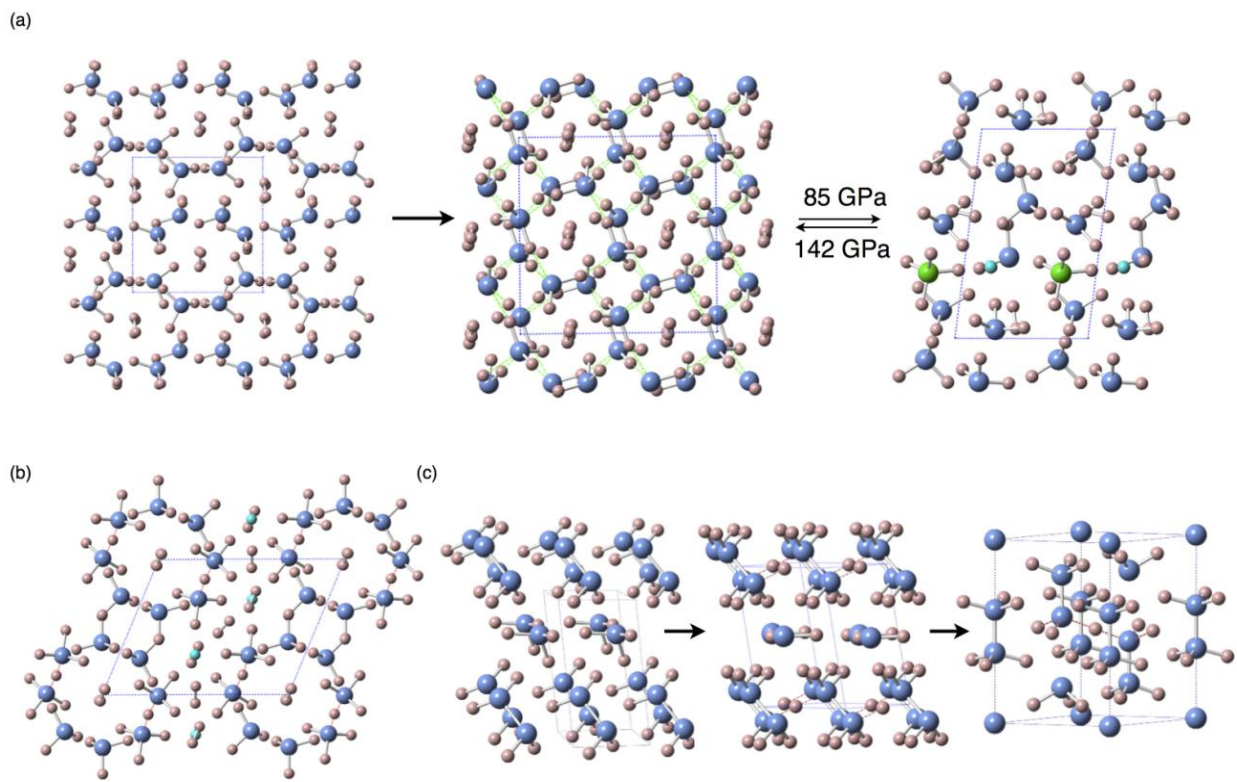

**Figure 4** The proposed structures for NH$_4$, NH$_5$ and N$_3$H$_7$. The small pink spheres indicate hydrogen atoms and the blue large spheres are nitrogen atoms. The nitrogen atom in NH$_4$ cation and the H$^-$ anion are noted with green and aqua spheres, respectively. (a) Phase transition sequence from host-guest $Pc$→host-guest $C2/c$↔Partially ionic $P1$-NH$_4$ phases. In host-guest structure of $C2/c$-NH$_4$, the hydrogen molecules are captured in the channels formed by NH$_3$ molecules. In the partially ionic $P1$-NH$_4$ structure, the NH$_4^+$ cation is close to the H$^-$ anion. (b) The ionic $C2/c$ NH$_5$ phase, with symmetric hydrogen bonds in [H$_3$N ⋯H⋯ NH$_3$]$^+$ units and H$^-$ anions. (c) Phase transition sequence molecular $P1$→ionic $C2$→ionic $P$-3$m$1 N$_3$H$_3$.

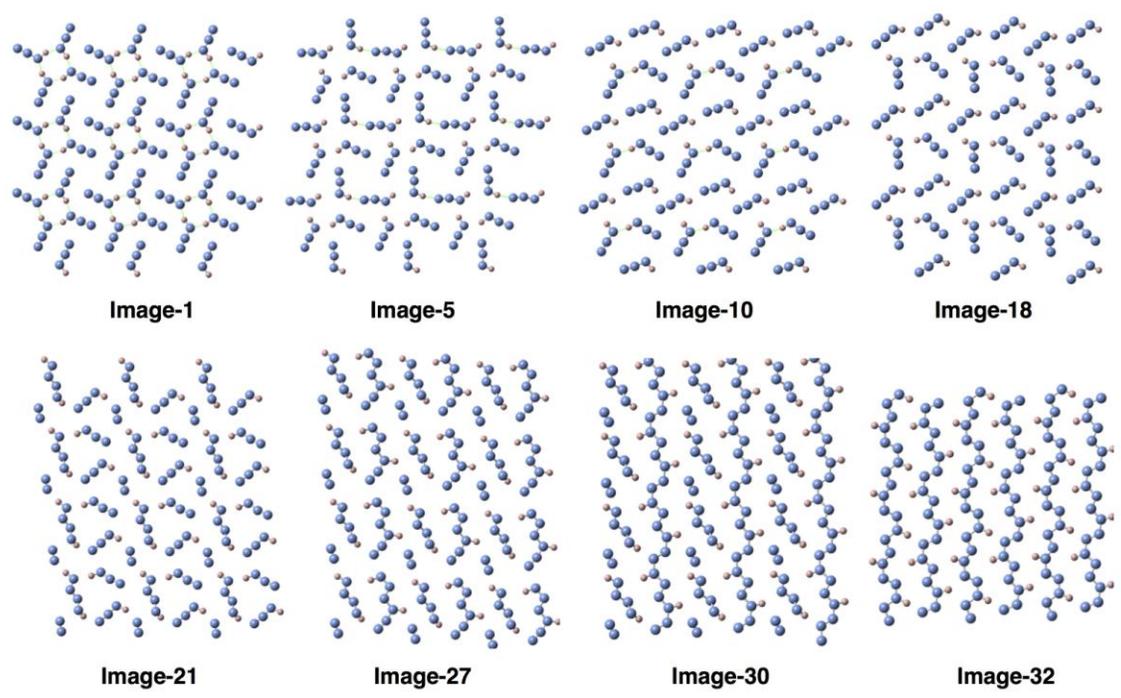

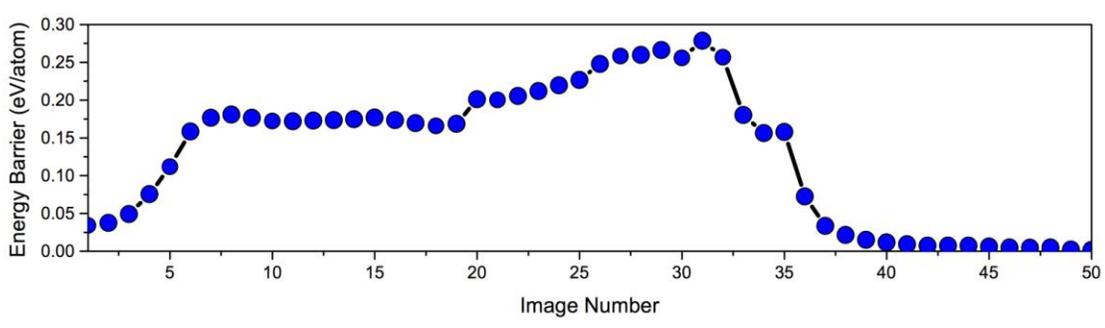

**Figure 5** Mechanism and energy barrier of hydrazoic acid to $P2_1/c$ $N_3H$ phase transition revealed by the VC-NEB method. A unit cell with 32 atoms was used during the pathway calculation. Only one layer of $N_3H$ structures during the phase transition are shown at specific images.